# Demonstration of Near-Infrared Negative-Index Materials

Shuang Zhang[1], Wenjun Fan[1], N. C. Panoiu[2], K. J. Malloy[1], R. M. Osgood[2] and S. R. J. Brueck[2§]

1. Center for High Technology Materials and Department of Electrical and Computer Engineering, University of New Mexico, Albuquerque, NM 87106

2. Department of Applied Physics and Applied Mathematics, Columbia University, New York, NY 10027

Metal-based negative refractive index materials have been extensively studied in the microwave region. However, negative-index metamaterials have not been realized at near-IR or visible frequencies due to difficulties of fabrication and to the generally poorer optical properties of metals at these wavelengths. In this paper, we report the first fabrication and experimental verification of a transversely structured metal-dielectric-metal multilayer exhibiting a negative refractive index around 2 μm. Both the amplitude and the phase of the transmission and reflection were measured experimentally, and are in good agreement with a rigorous coupled wave analysis.
PACS: 78.20.Ci; 42.25.Bs; 42.79.Dj

Negative refractive-index materials are of great interest for a variety of potential applications.[1] Because natural negative refractive index materials do not exist, artificial structures have been proposed and fabricated that exhibit an effective negative index over limited frequency ranges.[2] The two principal approaches to the realization of negative refraction are metamaterials and photonic crystals. Metamaterials typically use metallic structures to provide a negative permittivity and use resonant structures (inductor-capacitor tank circuits) with a scale much smaller than the wavelength to provide a negative permeability leading to negative refraction, while photonic crystals exhibit negative refraction as a consequence of band-folding effects. In the microwave region, negative index materials have been demonstrated using both approaches, while in the visible spectral region, negative refraction has been recently predicted.[3] In recent work, magnetically resonant structures exhibiting negative permeability have been demonstrated in the mid-infrared.[4,5] Despite theoretical studies and numerical modeling,[6,7] demonstration of negative refraction at near-infrared (near-IR) and visible wavelengths is as yet missing and, therefore, the experimental demonstration of a negative refractive index around 2 μm presented here represents an important milestone.

---

§ Email: brueck@chtm.unm.edu

Extension of metamaterials based on split ring resonators[8] to near-IR and visible wavelengths necessarily involves complicated and often difficult fabrication for the nanoscale metallic structures used to generate the requisite resonances. On the other hand, much of the photonic crystal literature has focused on all-dielectric structures because of their low-loss characteristics. Recent related work has used periodic metallic structures to couple incident radiation to surface plasma waves giving enhanced optical transmission through arrays of sub-wavelength holes in a metal film.[9,10] In this work, a hybrid approach is introduced which uses a pair of metal layers separated by a dielectric to provide resonant interactions (e. g. distributed inductance/capacitance) along with a periodic array of holes through the film stack to facilitate interaction with the surface plasma waves of the composite structure.

The structure consists of a glass substrate with two metallic films (30-nm thick Au) separated by a dielectric layer (60-nm thick $Al_2O_3$) with a 2-dimensional, square periodic array of circular holes (period 838 nm; hole diameter ~ 360 nm) perforating the entire multi-layer structure. A schematic and a SEM picture of the fabricated structure are shown in Fig. 1. As illustrated in Fig. 1(a), the structure can be roughly divided into two functional parts. The incident magnetic field interacts primarily with the hatched regions. The top and bottom metal films form a loop (inductor) and, where the induced current is interrupted by the hole, along with the capacitive coupling between the two films give a resonant response very similar to that recently reported for Au "staple" structures.[5] In the vicinity of the resonance, this tank circuit provides a magnetization field opposite to that of the incident wave and, therefore, a reduced permeability. The dark regions dominate the electric field response forming a wire grid polarizer that cancels the field in the metal and results in a negative permittivity.

The fabrication procedure is as follows. First, a thick layer of anti-reflection coating (ARC) and negative photo-resist (PR) is spun onto a glass substrate. Interferometric lithography with 355-nm UV source (third-harmonic of a YAG laser) is used to define a 2D array of holes in the PR at a pitch of 838 nm.[11] Titanium is then deposited on the wafer by e-beam evaporation, followed by a lift-off process to remove the PR and leave an array of Ti dots on top of the ARC. Reactive ion etching (RIE) with oxygen gas is used to etch through the ARC layer using the Ti dots as a selective etch mask, followed by three evaporations: 30 nm of Au, 60 nm of $Al_2O_3$, (the refractive index is about 1.65), and 30 nm of Au. Finally, a second lift-off process is carried out to remove the remaining ARC forming the final structure. The overall sample size was ~ 1 cm$^2$, providing a large area for measurements of the optical properties. This large-area nanoscale fabrication is a significant advantage of the parallel interferometric lithography patterning approach as compared with serial patterning techniques such as e-beam lithography.

In order to unambiguously determine the effective propagation constant (and hence refractive index) of this composite structure, it is necessary to measure both the amplitude and phase of the film transmission and reflection. The phase information is not available from measurements on a blanket metamaterial film (sample A). Two "metamaterial phase masks" (Fig. 2) were fabricated to allow interferometric measurement of this phase information. The first, sample B, consisted of stripes of the composite multi-layer patterned structure alternating with open area stripes at a pitch of 16 μm and was used for transmission phase measurements. For the reflection phase measurements (sample C), the blank areas are replaced with Au reflecting stripes atop the metamaterials at a finer pitch of 4.7 μm.

The transmission and reflectance measurements were performed using a Fourier transform infrared spectroscopy (FTIR). The transmission measurement is carried out at normal incidence to the wafer and the transmission through a clean BK7 glass wafer was used as the background spectrum to which the experimental spectra are normalized. For reflectance, the incident beam is incident on the wafer at 11° from the surface normal. For both measurements, unpolarized incident light is used because the structure is symmetric with respect to polarization at near normal incidence. The periods of the phase-masks were chosen so that only the zero-order transmission (reflection) from the samples is collected and the higher order diffracted light is outside the numerical aperture of the FTIR optical system. This provides a near zero-path length interferometric measurement of the phase of the transmission (reflection) coefficient referenced to the blank (metal) regions, even using the FTIR broadband source.

The transmission spectra of samples A and B are shown in Fig. 3(a) along with the rigorous coupled wave analysis (RCWA) modeling results (Fig. 3(b)) for sample A. The reflectance spectra of samples A and C are shown in Fig. 3(c) and the RCWA modeling for sample A are shown in Fig. 3(d). RCWA is a widely used algorithm for analysis of scattering electromagnetic waves from periodic structures.[12,13] The structure was modeled as an array of square holes (330 nm side dimension), which converged computationally faster than a model using round holes.

For sample A, the sharp structures in the transmittance and reflectance curves at about 1.3 µm correspond to the surface plasma wave at the glass:Au interface. The longer wavelength resonance at about 2 µm is attributed to the L-C circuit resonance described above in connection with the magnetic response. The multi-layer film stack also exhibits a surface plasma wave resonance in this spectral vicinity associated with the $Al_2O_3$:Au surface plasma wave shifted to longer wavelength by coupling between the adjacent $Al_2O_3$:Au interfaces. The connection between these interrelated phenomena will be discussed in more detail in a future publication. There is a dip in the transmission of sample B around 2 µm, which means that the light going through the two regions interferes destructively, so there must be a significant variation in the phase of the light transmitted through the metamaterial near that wavelength.

A simple Drude model is used for the dielectric constant of gold, $\varepsilon(\omega) = 1 - \omega_p^2/[\omega(\omega+i\omega_c)]$, where $\omega_p$ = 1.37x$10^{16}$ Hz is the plasma frequency and $\omega_c$ = 4.08x$10^{13}$ Hz is the scattering frequency for bulk gold.[14] The thin metallic film (30 nm) in our device has more surface and boundary scattering than bulk material and, therefore, likely exhibits a higher scattering frequency. Thus, the transmission was modeled with scattering frequencies of 1, 2 and 3 times that of bulk gold as shown in Fig. 3(b). The experiment results are best fit by the model with a scattering frequency three times that of bulk gold; this scattering frequency was used in the following analysis.

The transmitted intensity for sample A is $T_A$ and the complex coefficient of transmission is $t_A = \sqrt{T_A} e^{i\varphi_A}$, while the transmission through the sample B is $T_B$. The relative areas of the metamaterial and blank stripes are $g_1$ and $g_2$, respectively, where $g_1 + g_2 = 1$. The zero order transmission of the phase mask (sample B) is

$$T_B = | g_1 t_A + g_2 t_2 e^{jk_0 D} |^2 = g_1^2 T_A + g_2^2 T_2 + 2g_1 g_2 \sqrt{T_A T_2} \cos(\varphi_A - k_0 D) . \quad (1)$$

$T_2$ is the transmission at the air-glass interface (taken as unity because the spectra are normalized to background measurements for a glass window), $k_0 = 2\pi/\lambda$, and $D$ is the thickness of the negative index material. There is a remaining ambiguity in the sign of the argument of the cosine. Analytically calculating the transmission phase for an unpatterned multilayer structure at long wavelengths ~ 2.5 µm, well beyond the resonance frequency, and comparing it with the experimental result uniquely resolves this ambiguity. This approach also resolves another problem associated with the sensitivity of the analyzed transmission phase $\varphi_A$ to the relative ratio $g_1$ and $g_2$ resulting from the very small value of $T_A$ at long wavelengths; the analytically calculated phase at long wavelength beyond the resonance is very stable against small variations of the geometry and of the gold parameters and can be used to narrow the uncertainty in the SEM measurement of $g_1$ and $g_2$. The values used in the fitting were well within the measurement uncertainty. For the extraction of the reflectivity phase, $T_A$ and $T_B$ are replaced with $R_A$ and $R_C$ in Eq. (1) and the argument of the cosine by $\phi_A + 2k_0 d - \delta$, where $\phi_A$ is the reflectance phase angle of the negative index material, $d$ is the height difference between the surfaces of the metamaterial and the gold; and $\delta$ is the optical phase change on reflection from the gold. Because the metal is not a perfect conductor, $\delta$ is not exactly $-\pi$, but is evaluated from the Drude model and is found to be insensitive to the variation of the scattering frequency used. We also calculated $\delta$ at a number of wavelengths from the measured refractive index of gold[14] and verified that the Drude model provides a good approximation.

Using Eq. (1), the transmission and reflectance phase information is extracted as shown in Fig. 4. The measured phase for both transmission and reflectance are in good agreement with the modeling results. Finally, we can use the transmission and reflectance amplitude and phase measurements to evaluate the metamaterial refractive index ($n = \sqrt{\varepsilon\mu}$) and impedance ($\varsigma = \sqrt{\mu/\varepsilon}$). The appropriate formulae, modified to account for the glass output medium ($n_g$),[15] are:

$$\cos(nkd_m) = \frac{1 + n_g r^2 - t^2}{[n_g(1+r) + 1 - r]t}, \tag{2}$$

$$\varsigma = \frac{i[(1+r)/t - \cos(nkd_m)]}{n_g \sin(nkd_m)} \tag{3}$$

where $d_m$ is the thickness of the metamaterial film stack.

Using Eqs. (2) and (3), the complex value of the effective refractive index of the structures was evaluated solely from the experimental data. The results from both measurement and modeling are shown in Fig. 5(a) and 5(b), respectively. The real parts are negative over a range of wavelengths around 2 µm, where the imaginary part undergoes a strong modulation. The minimum value of the real part is about -2, while the imaginary part is larger than 3, which means that the negative index material exhibits significant loss associated with electron scattering in the thin metal films.

These results are the first experimental demonstration of a metal-dielectric negative index material at near-infrared wavelengths. From the results for $n$ and $\zeta$, the effective permeability $\mu_{eff}$ and permittivity $\varepsilon_{eff}$ can be evaluated (not shown). The real part of $\mu_{eff}$ shows a strong modulation and the imaginary part has a peak characteristic of a strong magnetic resonance in the wavelength region where $n < 0$. However, the minimum value of $\mu_{eff}$ is not negative (for both the measurement and the modeling results) as a result of the large scattering loss. The refractive index is expressed in terms of the permittivity and permeability as $n = \sqrt{(\varepsilon_1\mu_1 - \varepsilon_2\mu_2) + i(\varepsilon_1\mu_2 + \varepsilon_2\mu_1)}$, where $\varepsilon = \varepsilon_1 + i\varepsilon_2$ and $\mu = \mu_1 + i\mu_2$. To achieve a negative $n$ with $\varepsilon_1 < 0$ and $\mu_1 > 0$, $-\varepsilon_1\mu_2$ must be larger than $\varepsilon_2\mu_1$, as is the case near the metamaterial resonance.

In summary, the first metal-dielectric negative-index metamaterial at 2 μm, a wavelength about $10^4\times$ smaller than previously reported, has been demonstrated. This near-IR metamaterial has a high loss coefficient that could be improved using lower-loss metallic films or compensated for by adding a gain material between the two metallic layers. The achievement of negative refractive index materials at near-IR and optical frequencies should lead to increased applications and will allow more detailed investigation of the interesting properties of NIM since the transverse sample scale is ~$10^4$ wavelengths, making edge effects insignificant. Additionally, this work provides a new direction for the design and fabrication of negative refractive index materials and highlights the advantages of large-area interferometric patterning techniques for nanophotonic structures.

**Acknowledgement:** The UNM portion of this work was supported by the ARO/MURI in "Deep Subwavelength Optical Nanolithography" and by the DARPA University Photonics Research Program.

Figure Captions:

Fig. 1. : (Color online) (a) Schematic of the multi-layer structure consisting of an $Al_2O_3$ dielectric layer between two Au films perforated with a square array of holes (838-nm pitch; 360-nm diameter) atop a glass substrate. For the specific polarization, the active regions for the electric (dark regions) and magnetic (hatched regions) responses are indicated. (b) SEM picture of the fabricated structure.

Fig. 2: Top, schematic of sample B, the phase mask for transmission phase measurement. The measured fill factors $g_1$ and $g_2$ are 0.52 and 0.48, respectively and $D \sim 120$ nm. Bottom, schematic of sample C, the phase mask for reflection phase measurement. The measured fill factor $g_1$ and $g_2$ are 0.65 and 0.35 respectively, and $d \sim 215$ nm.

Fig. 3: (Color online) a) The measured transmission of samples A and B. b) RCWA modeling results for transmission of sample A with different scattering loss. Black, red and green curves are for a scattering frequency 1, 2 and 3 times that of bulk Au. Inset shows a comparison of the 3x curve with measurement across the 1.8- to 2.2-µm region. c) Measured reflection of samples A and C. d) RCWA simulation results for reflection of sample A with scattering frequency as a parameter.

Fig. 4. The measured (black) and modeled (gray) phase angle of transmission (top) and reflection (bottom) for sample A.

Fig. 5: The effective refractive index extracted from measurement (a) and from modeling (b) showing a resonance and a negative real part at ~ 2.0 μm.

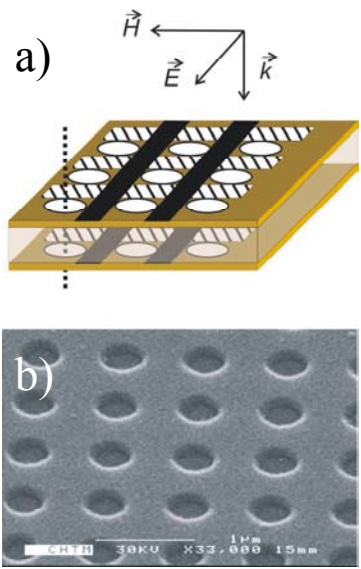

Fig. 1

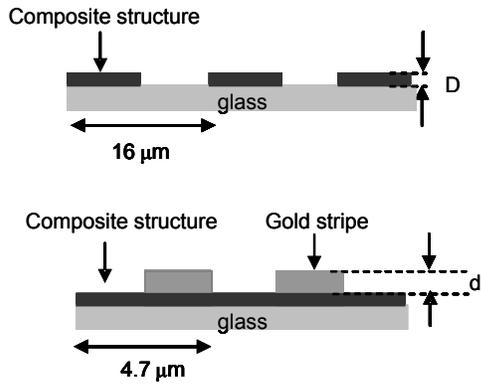

Fig. 2

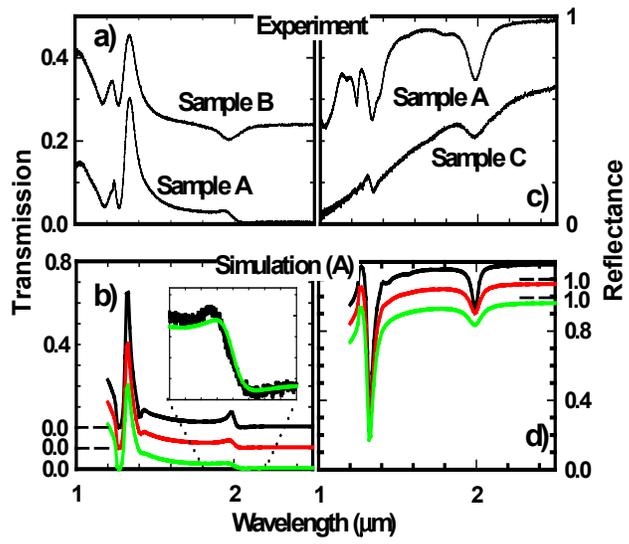

Fig. 3

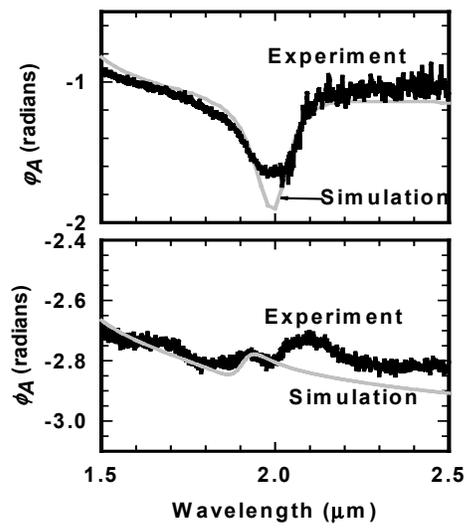

Fig. 4

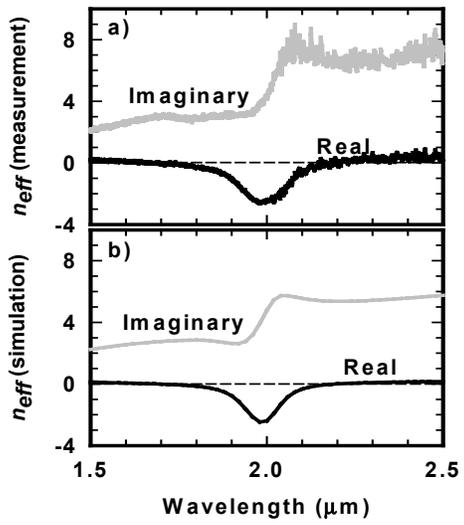

Fig. 5